# Turbulent convection and high-frequency internal wave details in 1-m shallow waters

**by Hans van Haren**

Royal Netherlands Institute for Sea Research (NIOZ) and Utrecht University, P.O. Box 59, 1790 AB Den Burg, the Netherlands.
e-mail: hans.van.haren@nioz.nl

*Abstract*

Vertically 0.042-m-spaced moored high-resolution temperature sensors are used for detailed internal wave-turbulence monitoring near Texel North Sea and Wadden Sea beaches on calm summer days. In the maximum 2 m deep waters irregular internal waves are observed supported by the density stratification during day-times' warming in early summer, outside the breaking zone of <0.2 m surface wind waves. Internal-wave-induced convective overturning near the surface and shear-driven turbulence near the bottom are observed in addition to near-bottom convective overturning due to heating from below. Small turbulent overturns have durations of 5-20 s, close to the surface wave period and about one-third to one-tenth of the shortest internal wave period. The largest turbulence dissipation rates are estimated to be of the same order of magnitude as found above deep-ocean seamounts, while overturning scales are observed 100 times smaller. The turbulence inertial subrange is observed to link between the internal and surface wave spectral bands.

Day-time solar heating from above stores large amounts of potential energy into the ocean providing a stable density stratification. In principle, stable stratification reduces mechanical vertical turbulent exchange, although seldom down to the level of molecular diffusion. Two contrasting mechanisms are thought to enhance turbulence in a stratified environment. The stratification supports internal waves 'IW' that may deform it by straining and vertical current shear. During night-time cooling, convective overturning may be found near the surface (Brainerd and Gregg, 1993). Both mechanisms pump heat down and nutrients up. The amounts of turbulence thus generated vary upon the efficiency of the mixing; convection, or 'Rayleigh-Taylor instability RTi', being generally more efficient than shear, or 'Kelvin-Helmholtz instability KHi', see, e.g., (Yabe et al., 1991; Dalziel, 1993). Generally in stratified waters, turbulent overturns are well separated from internal waves as the former last shorter and the latter longer than the buoyancy time-scale. The amounts of turbulence thus generated may be contrasted with that by the breaking of surface wind waves 'SW'.

In shallow, less than a few meters deep, sea areas like atolls, estuaries or near beaches, turbulence may also be generated by friction of sheared flow over the bottom (Ekman, 1905; Lamb, 1945), or by the convective heating from corals and the sediment (Jimenez et al., 2008; de la Fuente, 2014). It remains to be investigated whether these processes generate IW.

Heating from the atmosphere does not directly force IW, as less dense warm water initially spreads like a pancake over colder water forming a stable stratification. To generate IW this stratification has to be set into motion, either by frictional flow over the surface sucking up colder water (Ozen et al., 2006), or by (wave) flow over small bottom topography (Bell, 1975). Near a beach for example, IW are expected to be forced by flow over sand-banks (van Haren et al., 2012). IW-amplitudes are generally much larger than SW's and their periods O(50-1000 s) are much longer than the 5-10 s of SW, because of the weaker restoring force of reduced gravity.

In the present paper we are interested in quantifying turbulence processes using high-resolution instrumentation in shallow waters near beaches under calm atmospheric and day-time-heated stratified sea conditions to question: What are the dominant turbulent mixing processes? What are their dominant scales and appearances? The results may be portable to other near-surface ocean areas to scale up, including atoll lagoons, and to laboratory turbulence studies to scale down. After all, the ocean including the 1-m scales near beaches are still high bulk Reynolds number $>10^4$ areas.

*Materials and methods*

During the early summers of 2010 and 2011 instruments for IW-turbulence studies were on stand-by to be taken out the Dutch island of Texel beaches on short notice whenever conditions were right. These conditions implied SW (wind including swell) heights <0.2 m top-crest, sufficient daytime solar heating, Low Water 'LW' tide in early morning and late afternoon for positioning and recovering instrumentation. The primary instrumentation consisted of a wooden pole (Fig. 1) holding several tens of 'NIOZ4' self-contained high-resolution temperature (T) sensors at 0.042 m vertical intervals that sampled at a rate of 2 Hz with noise level of <0.0001°C and a precision of <0.0005°C (van Haren et al., 2009 for its predecessor NIOZ3 with similar characteristics). Around time of HW, the pole was fixed in about H = 0.5 m water depth to the sandy bottom using two concrete slabs, weighing 40 kg each (in air). Distance to shore, high-water 'HW' mark, was approximately 100 m. The tidal range varied between 1.5 and 2 m. Meteorological data were available from airport Den Helder, about 15 km southward from both sites investigated. SW-amplitudes and periods were estimated from the air-water surface passing the T-sensors.

In 2010, 36 T-sensors were mounted with the lowest at 0.13 m above the bottom 'mab'. Two sensors failed, including the lowest. The pole was moored at 53° 02.944′N, 04° 42.800′E, near Texel North Sea 'NS' beach-pole 13 (Fig. 1). In 2011, 52 T-sensors were mounted with the lowest at 0.048 mab. None failed. In addition, a single Sea-Bird Electronics SBE37 self-contained CTD was attached around 0.75 mab. The pole was moored at 53° 01.744′N, 04° 49.208′E, near Texel Wadden Sea 'WS' beach 'Ceres'. The NS is connected to the open ocean with relatively large SW-action. The WS is an inland tidal flat sea with closer connections to fresh water outflow.

The moored T-sensor data are used as a conservative estimate for dynamically more important density ($\rho$) variations, using standard relation, $\delta\rho = -\alpha\delta T$, $\alpha = 0.23$ kg m$^{-3}$ °C$^{-1}$ denoting the apparent thermal expansion coefficient under local conditions. Compressibility effects are negligible in the present data. Lacking salinity (S) data in 2010 it is assumed that only temperature contributes to $\delta\rho$ for the NS-beach observations. This assumption is justified as differential horizontal advection reinforces solar heating stratification, because in summer the fresher WS is warmer than the saltier NS (van Aken, 2008; van Haren, 2010). Thus, stable density stratification may occur with relatively cold above warm water, but not with relatively salty above fresh water. Hence, the above density-temperature relationship is a conservative one. This is confirmed in 2011, as the CTD gave $\alpha = 0.51$ kg m$^{-3}$ °C$^{-1}$ during the WS deployment.

With temperature as tracer for density variations turbulence can be quantified using the moored T-sensor data. Vertical turbulent kinetic energy dissipation rate ε, proportional to turbulent diapycnal flux, and eddy diffusivity $K_z$ are estimated by calculating overturning scales. These scales are obtained after reordering every time-step the potential density (temperature) profile ρ(z), which may contain inversions, into a stable monotonic profile ρ($z_s$) without inversions (Thorpe, 1977). After comparing raw and reordered profiles, displacements d = min(|z-$z_s$|)·sgn(z-$z_s$) are calculated necessary for generating the stable profile. Then,

$$\varepsilon = 0.64 d^2 N^3, \qquad (1)$$

where N denotes the buoyancy frequency computed from the reordered profile and the constant follows from empirically relating the overturning scale with the Ozmidov scale $L_O$ = 0.8d (Dillon, 1982), a mean coefficient value from many realizations. Using $K_z = \Gamma \varepsilon N^{-2}$ and a mean mixing efficiency coefficient for conversion of kinetic into potential energy of Γ = 0.2 for ocean observations (Osborn, 1980; Oakey, 1982; Gregg et al., 2018), we find,

$$K_z = 0.128 d^2 N. \qquad (2)$$

According to Thorpe (1977), results from (1) and (2) are only useful after averaging over the size of an overturn. In the following, 'sufficient' averaging is applied over at least vertical scales of the largest overturns and over at least buoyancy time scales to warrant a concise mixture of convective- and shear-induced turbulence, and to justify the use of the above mean coefficient values. Due to the small precision of the T-sensors, thresholds limit mean turbulence parameter values to $\langle\varepsilon\rangle_{thres} = O(10^{-12})$ $m^2 s^{-3}$ and to $\langle K_z\rangle_{thres} = O(10^{-6})$ $m^2 s^{-1}$ in weakly stratified waters (van Haren et al., 2015). In comparison with multiple shipborne shear- and temperature-variance microstructure profiling the present method yielded similar results to within a factor of two.

*Observations*

We focus on morning observations, when heating is up so that the air temperature $T_a > T_w$ the water temperature and near-surface waters are potentially stable for direct convective overturning. (Night-time convection indeed gave fully homogeneous, relatively cool waters due to convective overturning).

On 30 June 2010 near NS-beach, the pole was moored around sunrise, one hour after LW. Cloud-cover was relatively high (60-100%) the entire day, with some sunshine between days 180.3 and 180.4 UTC. A daily evaporation sum of 0.7 mm was calculated for these conditions following Lin (2007), implying a daily salting of 0.02 g kg$^{-1}$ m (van Haren et al., 2012). To balance this density contribution one requires about 0.1°C of warming; the observed T-differences of about 1°C were larger (Fig. 2a). Potentially stratification counteracting, free convection generating, ground water leakage is found of little importance. Easterly winds created only small SW near the pole, as the NS-beach is exposed to the West. SW-height was about 0.1 m (Fig. 2a) with 3-10 s periods.

Trains of well-stratified waters passed the T-sensors with vertical amplitudes varying between 0.2 and >1.0 m, all > SW-heights. T-variations have 'periods' of half an hour, and shorter with changes every 60-90 s. Local stratification has a mean buoyancy period of $T_N \approx 200$ s and a smallest buoyancy period of $T_{Nmax} \approx 60$ s when calculated across thin layers of <0.1 m thickness.

The largest turbulent overturns occur from the surface and push the stratification down (Fig. 2b). Turbulent overturns are $|d| > 0.5$ m and last up to $T_{Nmax}$, with the group of overturns around day 180.305 lasting as long as $T_N$ (Fig. 2c). This timescale of turbulent overturning well exceeds the classic initial RTi growth rates $(Ag2\pi/L)^{1/2} \approx 1$ s$^{-1}$ (Chandrasekhar, 1981) as, with acceleration of gravity g, for the present data Atwood number $A = 2\Delta\rho/\rho \approx 0.005$ and length-scale $L \approx 0.3$ estimated using phase speed 0.03 m s$^{-1}$ (van Haren et al., 2012). During this bursting, the resulting 1.5-m-vertical- and $T_N$-mean turbulence estimates are $[<\varepsilon>] = 4\pm2\times10^{-7}$ m$^2$ s$^{-3}$, $[<K_z>] = 7\pm3\times10^{-5}$ m$^2$ s$^{-1}$. Due to the small d-scales and high N, the former (~flux) value is relatively high, and the latter value relatively low, compared with deep-ocean turbulence exchange parameter values. It demonstrates that near-beach waters can be turbulent internally, also on calm days without SW-breaking. The period after day 180.315 shows a strong interface deepening by about 1 m in 400-500 s, the interface still about 0.2 m thick. As easterly winds were weak, the deepening is thought to be due to the warming and differential advection. The weaker vertical- and $T_N$-mean $[<\varepsilon>] = 2\pm1\times10^{-8}$ m$^2$ s$^{-3}$, $[<K_z>] = 3\pm1\times10^{-6}$ m$^2$ s$^{-1}$ turbulence values are generally occurring in very short—small-scale bursts of about 10 s and 0.3 m (see Details below). The spikes in Fig. 2d are thus not random noise.

This small-scale turbulence is spectrally characterized as follows (Fig. 3). Temperature variance is up to 100 times larger near the surface than near the bottom, depending on the frequency band. Near the surface, its peaks follow a $\sigma^{-5/3}$ '-5/3' slope with frequency $\sigma$ for the range $N < \sigma < \sigma_{SW}$. This suggests a shear-dominated turbulence inertial subrange (Tennekes and Lumley, 1972) or passive scalar (Cimatoribus and van Haren, 2015). This -5/3-range is distributed over a larger frequency range than commonly found in the ocean and linked to the SW-band here. The spectra's bases outside the peaks seem to slope with -1, mainly in the range $0.5N < \sigma < 2N_{max}$. This points at some, weaker, influence of convective turbulent overturning or active scalar, and, potentially only for $0.5N < \sigma < N$, linear internal waves as observed in the open ocean (van Haren and Gostiaux, 2009). The latter comparison is dubious for the present observations, given the irregular motions in Fig. 2 and upcoming details. With reference to the data of the T-sensor at 0.65 mab, the coherence (Fig. 3c) is only significant between directly neighbouring sensors and further away for $\sigma < N_{max}$ (2 neighbouring sensors) and $\sigma < N$ (4 neighbours). At higher frequencies, no significant coherence is found between directly neighbouring sensors.

**Details**

An apparent wind-shear sucking up cold-water penetrating the warmer near-surface layer is seen to consist of two (sets of) overturns (Fig. 4). The turbulence lasts about 40-50 s < $T_{Nmax}$ and is found to be a complex of smaller overturn(s) in larger one(s). The more or less singular event does not immediately correspond with a series of cusps as in Ozen et al. (2006). It is noted that wind speeds are low <4 m s$^{-1}$. Before and after the singular event multiple numerous small-scale overturning is observed of |d| = 0.1-0.15 m and lasting 5-20 s. These are observed throughout the water column.

Such small-scale overturning is also observed in other periods, although less numerous when more intense turbulence occurs above (e.g., Fig. 5). Here also an apparent 1 m upward cusp is seen, squeezed between two large turbulent downdraughts. Around 0.5 m the stratification is stronger than in Fig. 4, hence the potentially fewer small overturns, being forced down by the large overturning above. Mean values over the 1.5-m and 250-s ranges are [<$\varepsilon$>] > 10$^{-7}$ m$^2$ s$^{-3}$, [<$K_z$>] > 10$^{-4}$ m$^2$ s$^{-1}$, with largest values in the relatively cooler updraughts suggestive of shear-dominated turbulence.

However, relatively strong turbulent overturning is also observed in warm downdraughts, suggesting convection-dominated turbulence (Fig. 6), under conditions when principally $T_a$ > $T_w$, on the large scale. As before, stratification is pushed down and small turbulent overturning underneath is dampened. Small turbulent overturning reappears when main stratification moves up and weakens in thicker layering, around day 180.303. While the larger overturns last about 15-20 s, around day 180.3025, the small turbulent overturns last 4-5 s.

An example of large-scale overturning in an 'interior' relatively thick near-homogeneous layer is given in Fig. 7. It lasts about $T_{Nmax}$, while the local minimum buoyancy period $T_{Nmin}$ ≈ 1000 s within the particular layer. It is thus not likely associated with intruding partially salinity-compensated waters that can last >$T_{Nmin}$. The association with the thin-layer stratification motions is not directly obvious. These IW-motions vary at very high-frequency close to $\sigma_{SW}$, but they are highly irregular, more nonlinear. The latter may be due to complex wave-interactions, or, more likely here, to interior turbulence (higher-up) interactions with stratification.

**Inland-sea beach**

The connection between nonlinear IW and turbulence is also observed in 2011, near a WS-beach (Fig. 8). In this image, comparable to Fig. 2 with similar mean turbulence levels outside the periods with large near-surface levels, progressively larger overturns approach the size of the mean buoyancy period and about |d| = 0.3 m, >> SW. This shear-induced KHi, best visible in the S-shapes in Fig. 8b, comes with small secondary KHi along its edges as observed in estuaries by Geyer et al. (2010). Near the bottom, turbulent overturning is seen to affect IW-motions near the interface, around day 185.4 the time of HW flow reversal. This is due to the gradual warming of the water column in addition to and possibly driving the shear-flow. It is seen that for episodic periods of 20 s to local $T_N$, the smallest IW-timescales, the bottom is apparently warming the water from below hence initiating convective overturning thereby generating some high-frequency interfacial IW. It compares with modeling results on a very shallow <0.1 m deep salt-water lake (de la Fuente, 2014) where solar heated sediment warms the water from below. While the present total duration of overturning lasts just under local $T_{Nmin}$ = 1000 s, the near-bottom turbulence is not expected to be driven by bottom frictional shear flow. Such a flow would either generate turbulence during the general cooling phase only (Lorke et al., 2005; van Haren, 2010) or drive warmer waters over cooler near-bottom waters in rotational Ekman (1905) dynamics off a slope. Both are not observed here during the gradual warming with the largest episodic warming at the lowest sensor. In addition, the present observations show a reduction with time of the near-homogeneous near-bottom layer, instead of a thickening.

This results in a spectral overview that has a clearer $\sigma^{-5/3}$ slope of shear-induced turbulence for $N < \sigma < \sigma_{SW}$ and $\sigma^{-1}$ of convective turbulence for $N < \sigma < 2N_{max}$ (Fig. 9) in comparison with NS-data in Fig. 3. With respect to the T-sensor at 0.5 mab, coherence over more than one neighbouring sensor away is only found for $\sigma < N$, and, barely significant, in the SW-band (Fig. 9c).

As the -5/3 inertial subrange is observed up to the SW-range, small-scale energy containing turbulence has timescales as short as 3-5 s at least. (Turbulence dissipation scales are O(1 mm) and O(0.01 s) in the ocean). However, the small-scale overturns occur distributed in the interior of the depth-time range and have no vertical coherence (over scales down to 0.042 m), so that they are not directly associated with SW. This indirectly confirms the non-isotropy of stratified turbulence by a factor of at least two, providing aspect ratios of <0.5. The observations were never made during times and zones of breaking of SW, which are thus assumed to propagate more or less linearly. While the inertial subrange IW-induced-turbulence- and SW-bands directly associate in frequency, the transfer of energy between the two is not established.

*Discussion*

The present observations demonstrate that IW can occur near shallow beaches in open and enclosed seas, and presumably also in lakes. Any swimmer and paddler should be able to feel them on a calm day, provided $T_w$ > 15°C to avoid pain blocking perception (Kuhtz-Buschbeck et al., 2010). Although they are only observed during calm days outside SW-breakers their turbulent exchange is not negligible. With respect to open-ocean IW-turbulence, the spatial scales near the beach are O(1 m), 100 times smaller. In 2-m deep waters, the 0.1-1 m large IW-amplitudes occur in conjunction with turbulence overturns of similar sizes. While the dominant turbulence-generation process seems to be shear-induced KHi, judging from the spectra, convective RTi occur frequently in the background. The largest RTi are observed near the surface, hypothesized to be driven by differential advection deforming small-scale IW, but also in the interior after IW-straining creating near-homogeneous layers, and near the bottom by episodic heating from below. The RTi associate with IW-generation at the interfacial layers above or below. The precise mechanisms of IW-turbulence interaction, including the association with SW, requires further future theoretical modelling investigation extending works e.g. by Thorpe (2010) for continuous open-ocean stratification.

The resulting mean turbulence dissipation rates, which are proportional to turbulent fluxes, are estimated to be within one order of magnitude of those of the largest deep-ocean IW-breaking above seamounts: $\varepsilon = O(10^{-7})$ m$^2$ s$^{-3}$. The $\varepsilon$-value is 10-100 times larger than that of open-ocean 'linear' IW-breaking (e.g., Gregg, 1989). In contrast, mean turbulent diffusivities $O(10^{-5})$ m$^2$ s$^{-1}$ found near beaches are comparable to those from the open ocean and are thus two to three orders of magnitude smaller than those observed above seamounts (e.g., van Haren et al., 2015). This is due to the small length-scale in the shallow waters near beaches, in association with the strong stratification.

The turbulence dissipation rate level of night-time convection is of the same order of magnitude as modest SW-breaking at a beach (Brainerd and Gregg, 1993): Both are about 100 times larger than the mean dissipation rates estimated in the present data. However, short-term peaking of IW-breaking may develop turbulent flux levels up to those of weak-modest SW-breaking and near-surface night-time convection. As a result, under calm atmospheric and, day-time-heated, stratified sea conditions IW-turbulence may be important for vertical exchange in shallow waters, not only near a beach, but also in estuaries, atolls, and near the ocean surface, although in the latter heating from below will not contribute.

**Acknowledgments**

I thank M. Laan and L. Gostiaux for their collaboration in design and construction of NIOZ T-sensors. The sensors have been financed in part by NWO, the Netherlands Organization for Scientific Research.

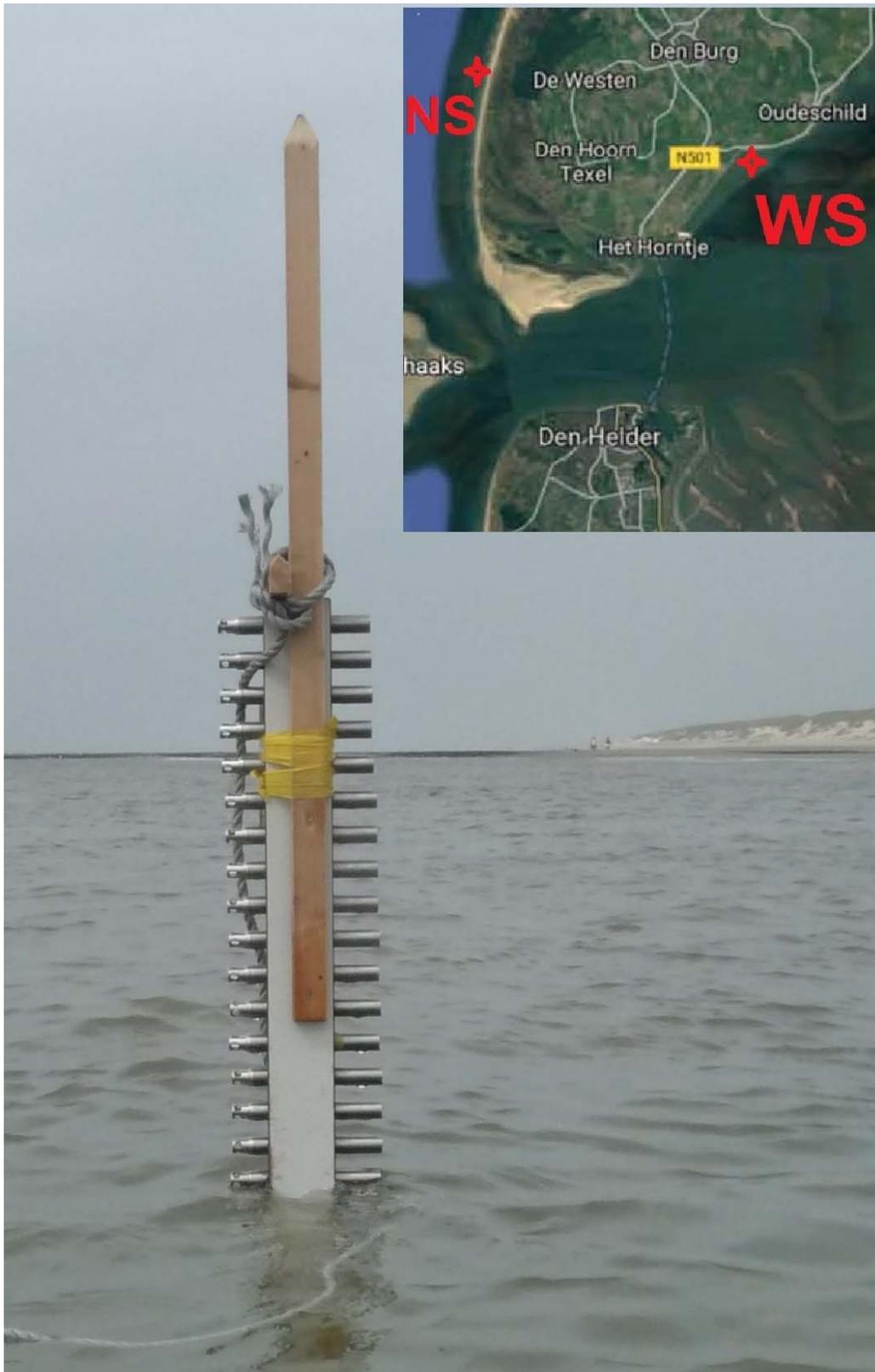

**Figure 1**. Instrument pole with upper sensor at 1.60 m above the bottom 'mab' near Texel North Sea 'NS' beach, around Low Water 'LW' on day 180.638 UTC, 2010. Insert shows the two mooring locations, including the 2011 Wadden Sea 'WS' beach.

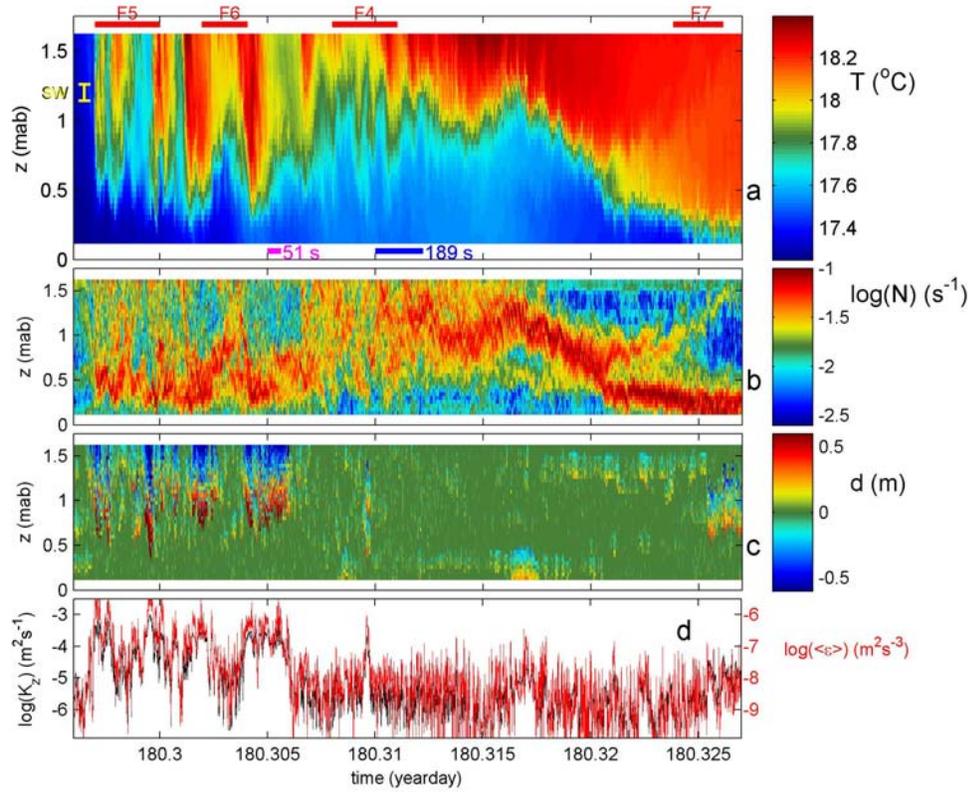

**Figure 2**. About 45 min overview of internal wave turbulence observations near Texel NS-beach on 30 June 2010, on average one hour (0.04 day) before HW (on day 180.3542) and one hour after the morning equality of air and water temperature (on day 180.261). (a) Depth-time series of moored T-observations. The vertical scale is 1.75 m with reference to the bottom. The horizontal red bars indicate magnifications in forthcoming Figures (letter F + number). The purple and blue horizontal bars indicate the minimum and mean buoyancy period for the panel, respectively. The yellow vertical bar on the left indicates the spread of surface wave 'SW' height, estimated from the water surface passing the sensors. (b) Logarithm of buoyancy frequency after reordering a. to stable profiles every time step and using $\delta\rho = -0.23\delta T$ kg m$^{-3}$ °C$^{-1}$ (see text). (c) Overturning displacements following comparison of a. with its reordered data. (d) Time series of logarithms of vertically averaged eddy diffusivity (black, scale to the left) and turbulence dissipation rate (red, scale to the right).

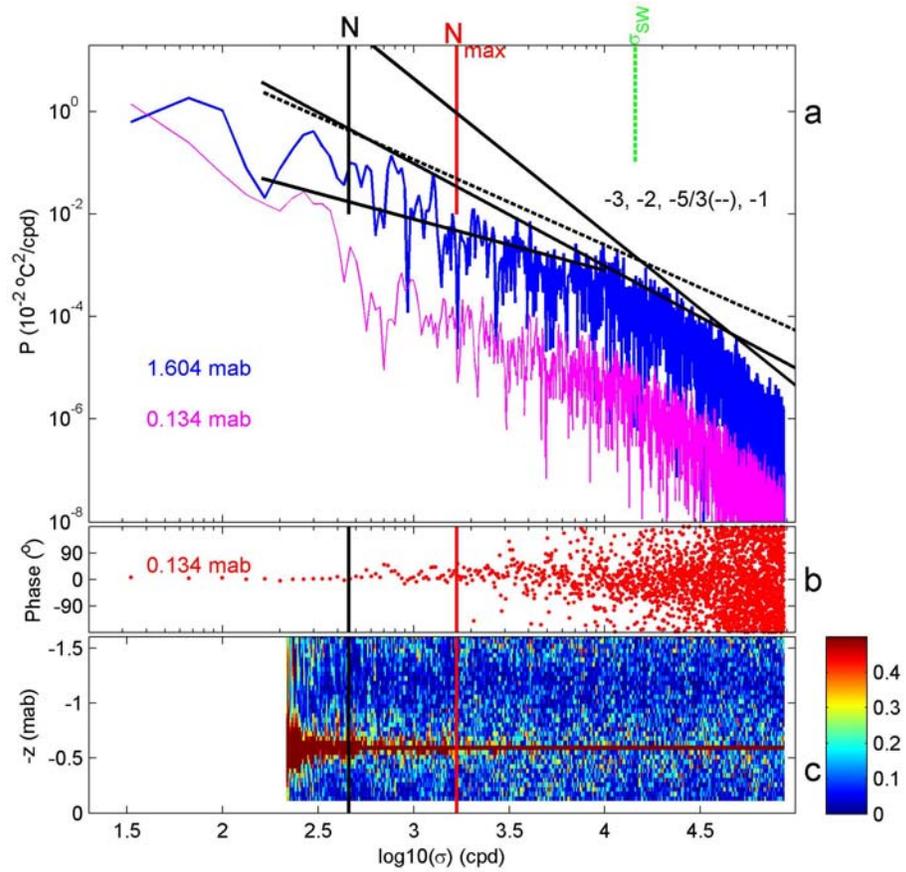

**Figure 3**. Spectral view of the data in Figure 2. (a) Weakly smoothed (3 dof, degrees of freedom) temperature variance for the upper- (blue) and lowermost (purple) levels. (b) Weakly smoothed (3 dof) phase for temperature correlation between the lower two sensors. (c) Depth-frequency series of moderately smoothed (20 dof) coherence between the sensor at the dark-red depth level and all other sensors. The 95% confidence threshold-level is at approximately 0.2 coherence.

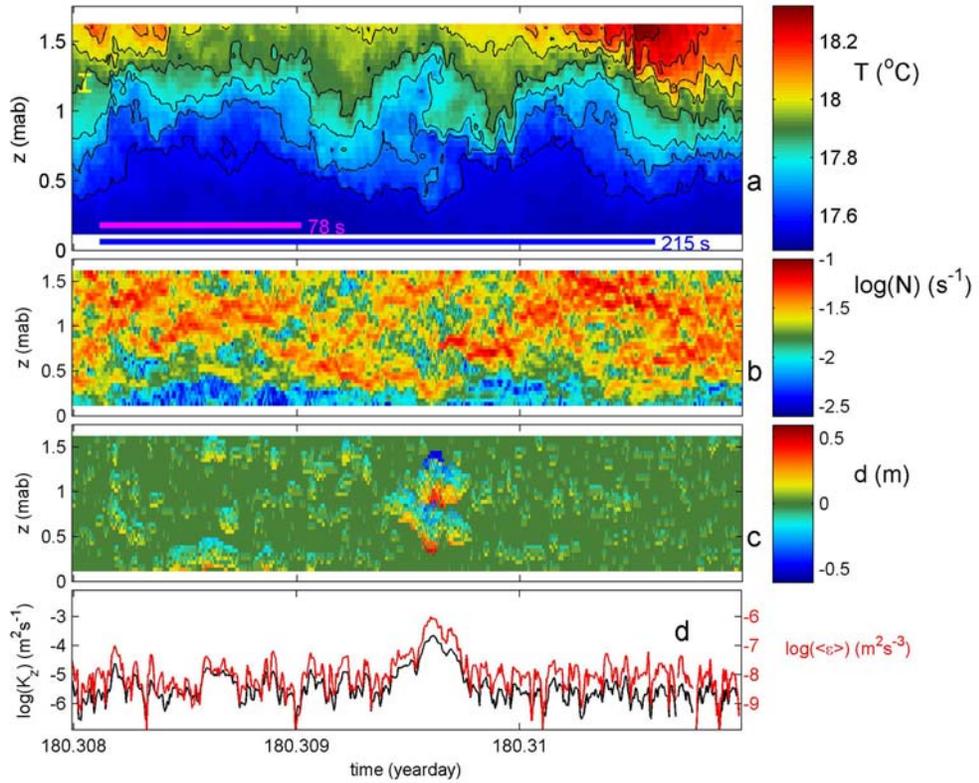

**Figure 4**. As Figure 2, but for a 4 min detail of double Kelvin-Helmholtz instability 'KHi' overturning (peak at day 180.3096) and numerous small-scale, weakly turbulent overturns causing a broken and broad interfacial layer. In a., the colour-scale is slightly different than in Fig. 2a and black contours are drawn every 0.1°C.

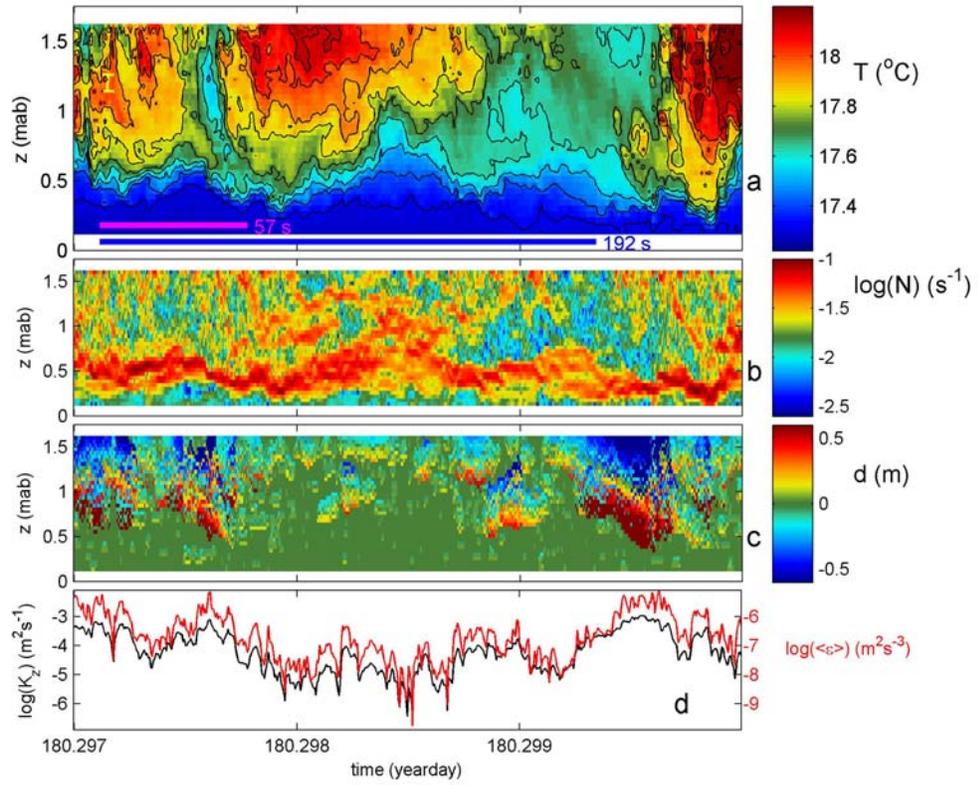

**Figure 5**. As Figure 4, but for a 4 min detail of convective overturning above an interface around 0.5 mab oscillating with the smallest internal wave period (purple bar) possible.

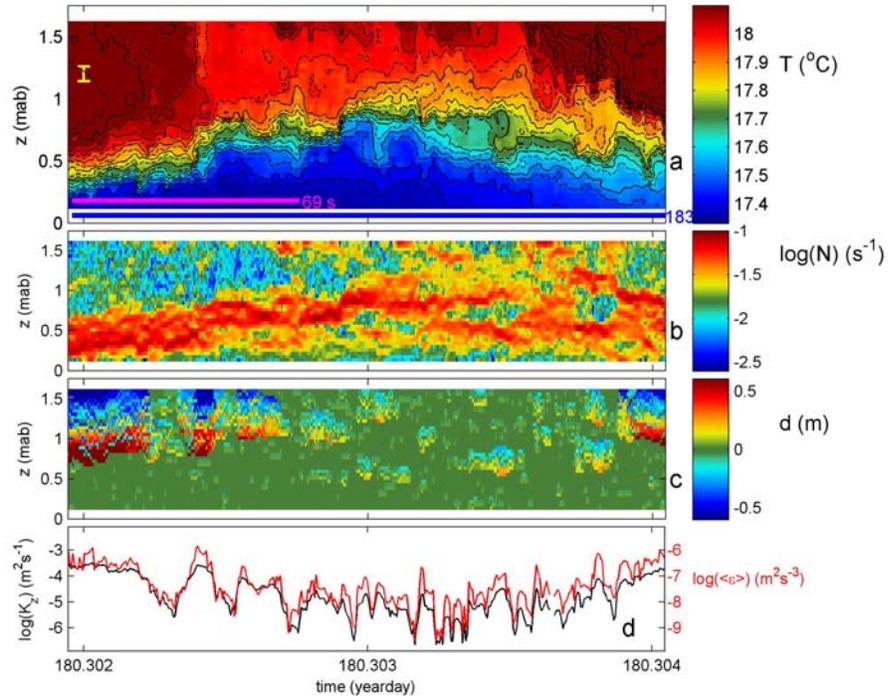

**Figure 6**. As Figure 4, but for a 3 min (one-mean-buoyancy-period) detail of convective overturning above an initially 0.25 m thick interface that slowly moves upward (by internal wave motion), and subsequently broadens and breaks up by multiple KHi like on day 180.303. In a., the dashed black contours indicate 0.05°C intervals.

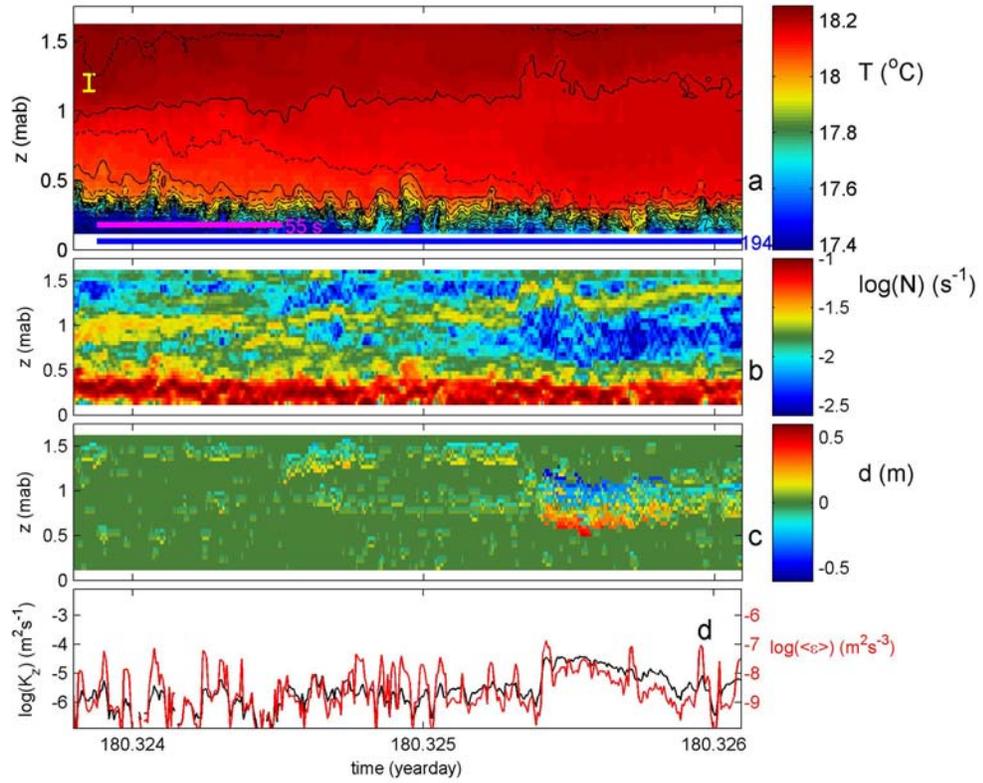

**Figure 7**. As Figure 6, but for a one-mean-buoyancy-period detail of 'internal convective overturning' after day 180.3254 around 0.9 mab above a 0.15 m thick interface close to the bottom, which shows a sequence of rapid KHi around day 180.325.

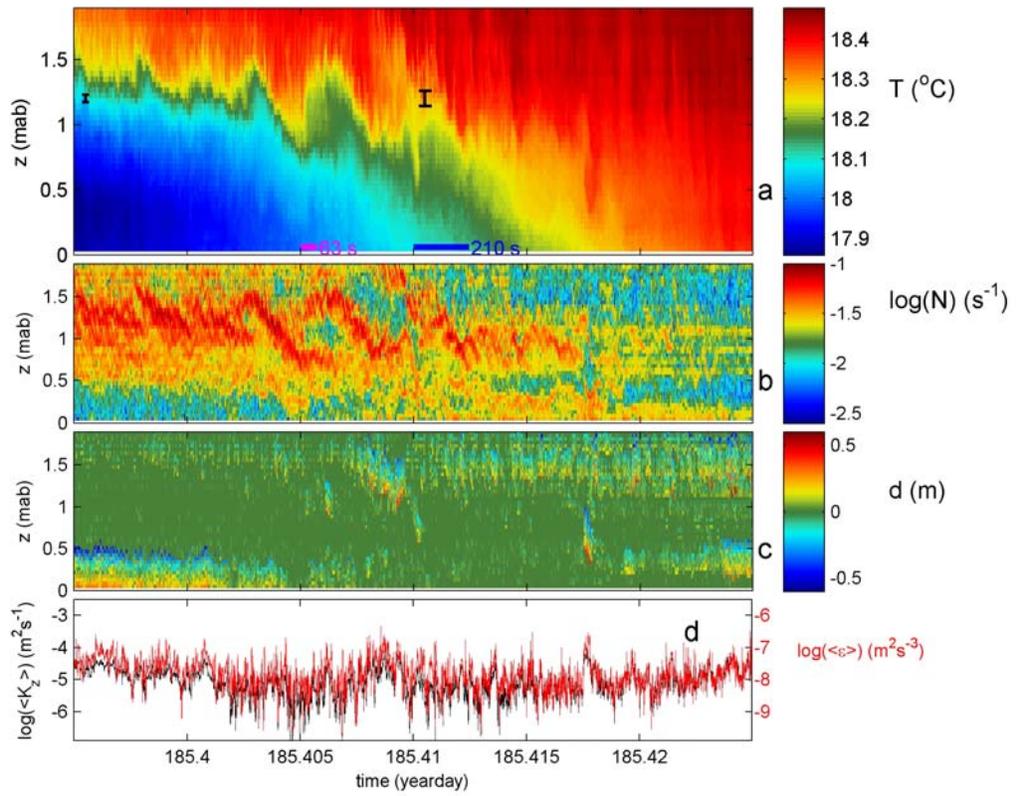

**Figure 8**. As Figure 2, but for a 45 min of data on 05 July 2011, off 'Ceres', a WS-beach with small wave action and larger salinity stratification. Due to stable salt stratification $\delta\rho = -0.51\delta T$ kg m$^{-3}$ °C$^{-1}$, as established from a single SBE37 CTD at 0.75 mab. The large S-shaped forms in b. are indicative of single KHi-overturning. HW was at day 185.4104 UTC; waves/wind/sunshine increase at day 185.41.

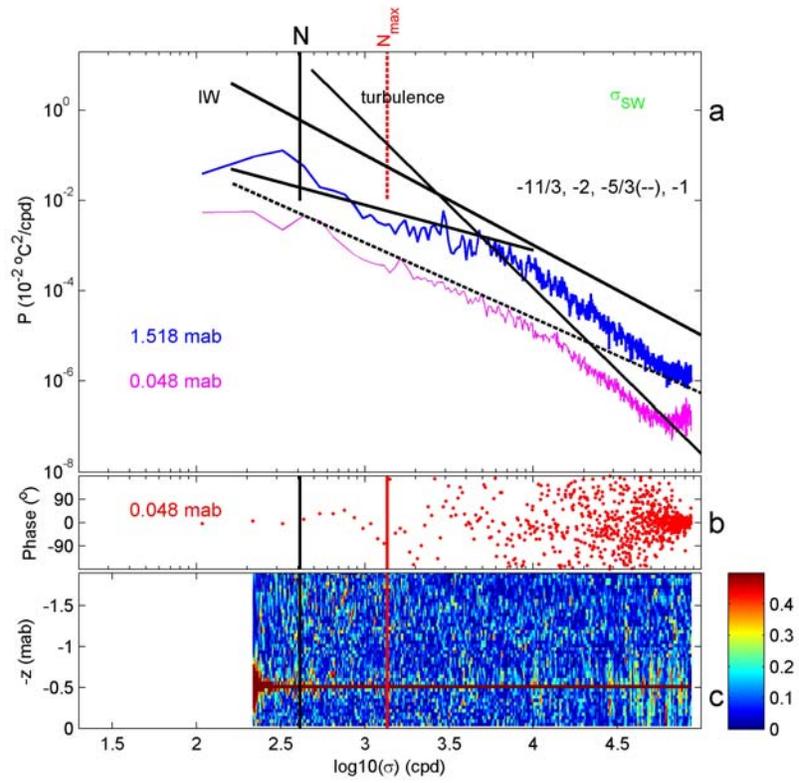

**Figure 9**. As Figure 3, but for data in Figure 8, with data in a. and b. moderately smoothed (20 dof) due to averaging over data from 7 sensors each.